\documentclass[twocolumn,showpacs,superscriptaddress,amsmath,amssymb]{revtex4-1}
\usepackage{graphicx}
\usepackage{epsfig}
\usepackage{dcolumn}
\usepackage{bm}
\usepackage{overpic}
\usepackage{color}
\usepackage{subfigure}
\usepackage{threeparttable}

\def \lambdac    {\Lambda_{c}}
\def \lambdacp   {\Lambda_{c}^{+}}
\def \lambdacm   {\bar{\Lambda}_{c}^{-}}

\def \pipi  {\pi^{+}\pi^{-}}

\def \qqbar {q\bar{q}}

\def \kpkm  {K^+K^-}

\def \ee   {e^+e^-}
\def \gev  {\mbox{GeV}}
\def \gevc {\mbox{GeV/$c$}}
\def \gevcc{\mbox{GeV/$c^2$}}
\def \mev  {\mbox{MeV}}
\def \mevcc{\mbox{MeV/$c^2$}}
\def \ipb  {\mbox{pb$^{-1}$}}
\def \vr   {V_{r}}
\def \vz   {V_{z}}
\def \BR   {\mathcal{B}}

\def \epem {e^+e^-}
\def \pppm {\pi^{+}\pi^{-}}
\def \piz  {\pi^0}
\def \pip  {\pi^+}
\def \pim  {\pi^-}

\def \costht {\cos\theta}

\def \eff  {\varepsilon}

\def \costht {\cos{\theta}}

\begin{document}
\title{\boldmath Evidence for the singly-Cabibbo-suppressed decay $\lambdacp\to p\eta$ and search for $\lambdacp\to p\piz$}
\author{
  M.~Ablikim$^{1}$, M.~N.~Achasov$^{9,e}$, S. ~Ahmed$^{14}$,
  X.~C.~Ai$^{1}$, O.~Albayrak$^{5}$, M.~Albrecht$^{4}$,
  D.~J.~Ambrose$^{45}$, A.~Amoroso$^{50A,50C}$, F.~F.~An$^{1}$,
  Q.~An$^{47,a}$
% ...
  \\
  \vspace{0.2cm}
  (BESIII Collaboration)\\
  \vspace{0.2cm} {\it
    $^{1}$ Institute of High Energy Physics, Beijing 100049, People's Republic of China\\
% ...
$^{4}$ Bochum Ruhr-University, D-44780 Bochum, Germany\\
    $^{5}$ Carnegie Mellon University, Pittsburgh, Pennsylvania 15213, USA\\
% ...
$^{9}$ G.I. Budker Institute of Nuclear Physics SB RAS (BINP), Novosibirsk 630090, Russia\\
% ...
$^{14}$ Helmholtz Institute Mainz, Johann-Joachim-Becher-Weg 45, D-55099 Mainz, Germany\\
% ...
$^{45}$ University of Rochester, Rochester, New York 14627, USA\\
% ...
$^{47}$ University of Science and Technology of China, Hefei 230026, People's Republic of China\\
% ...
$^{50}$ (A)University of Turin, I-10125, Turin, Italy; (B)University of Eastern Piedmont, I-15121, Alessandria, Italy; (C)INFN, I-10125, Turin, Italy\\
% ...
$^{a}$ Also at State Key Laboratory of Particle Detection and Electronics, Beijing 100049, Hefei 230026, People's Republic of China\\
% ...
$^{e}$ Also at the Novosibirsk State University, Novosibirsk, 630090, Russia\\
% ...
  }
}

\date{\today}
\begin{abstract}
  We study the singly-Cabibbo-suppressed decays $\lambdacp\to p\eta$ and $\lambdacp \to p\piz$ using $\lambdacp\lambdacm$ pairs produced by $e^+e^-$ collisions at a center-of-mass energy of $\sqrt{s} = 4.6\;\gev$.  The data sample was collected by the BESIII detector at the BEPCII collider and corresponds to an integrated luminosity of 567 $\ipb$.
   We find the first evidence for the decay $\lambdacp\to p \eta$ with a statistical significance of $4.2\sigma$ and measure its branching fraction to be $\BR(\lambdacp\to p\eta)= (1.24\pm0.28({\rm{stat.}})\pm0.10({\rm{syst.}}))\times10^{-3}$.
  No significant $\lambdacp \to p\piz$ signal is observed.  We set an upper limit on its branching fraction $\BR (\lambdacp \to p\piz) < 2.7 \times 10^{-4}$ at the 90\% confidence level.
\end{abstract}


\maketitle

  Weak decays of charmed baryons provide a unique testing ground for different theoretical models and approaches, $e.g.$ the quark model approach to non-leptonic charm decays and Heavy Quark Effective Theory ~\cite{1998YKohara, 1998Mikhail, 1997KKSharma, 1994TUppal, 1994PZenczykowsky, 1992JGKorner, 1996LLChau}.
  The charmed baryon ground state $\lambdacp$ was first observed in 1979~\cite{1980GSAbrams,1979AMCnops},
  but,
  compared to the rapid advances of charmed mesons,
  progress in the studies of the charmed baryons
  has been relatively slow due to a lack of experimental data and the additional difficulties of three constituent quarks in theoretical calculation.
  The accuracy of $\lambdacp$ branching fractions (BFs) has long been poor for the Cabibbo-favoured (CF) decays, and even worse, with uncertainties at the 40\% level, for the singly-Cabibbo-suppressed (SCS) decays~\cite{2016PDG}.
  As a consequence, it is neither possible to test the BFs predicted by different theoretical models, nor to determine the effects of final-state interactions (FSI).
  It is therefore essential to improve the accuracy of these BFs for $\lambdacp$ decays and to search for new decay modes.
  The absolute BFs of twelve $\lambdacp$ CF hadronic decay modes have been measured by the BESIII collaboration with much improved precision~\cite{2015MAblikimLambdac}.

  The SCS decays $\lambdacp\to p\eta$ and $p\piz$ have not yet been studied experimentally.
  These two decays proceed predominantly through internal $W$-emission and $W$-exchange diagrams, which are non-factorizable and not subject to color and helicity suppression in charmed baryon decay.
  Some theoretical models~\cite{1997KKSharma,1994TUppal,2003SLChen,2016lvcaidian}, predict the BFs of these two process under different assumptions (the flavor SU(3) symmetry, FSI) obtaining different results.
  Therefore, measurements of these BFs will help us to understand the underlying dynamics of charmed baryon decays and distinguish between the different models.
  Furthermore,  the ratio of BFs of these two decays, which is expected to be relatively insensitive to the values of input parameters in the theoretical calculation, is an excellent probe to distinguish between the different models.

  In this Letter, we present the first experimental investigations of the SCS decays $\lambdacp\to p\eta$ and $p\piz$.
  We use a data sample of $e^+e^-$ collisions at a center-of-mass (c.m.) energy of $\sqrt{s}=4.6\;\gev$~\cite{2016Gaoqing} with an integrated luminosity of 567 $\ipb$~\cite{2015MAblikimLumi} collected by the BESIII~\cite{2009MAblikimDet} detector at the BEPCII~\cite{1994JZBei} collider.
  Taking advantage of the excellent BESIII detector performance and the clean environment just above the mass threshold to produce $\lambdacp\lambdacm$ pairs, 
  a single-tag method, ($i.e.$, reconstruction of only one $\lambdac$ in the $\lambdacp\lambdacm$ pairs) is used to increase the detection efficiency and acquire more $\lambdac$ candidates.
  Throughout the text, the charge conjugate states are always implied unless mentioned explicitly.

  $\uchyph=0$BESIII~\cite{2009MAblikimDet} is a cylindrical spectrometer, 
  consisting of a small-celled, Helium-based main drift chamber (MDC), a plastic scintillator Time-of-Flight system (TOF), a $\mbox{CsI}$(Tl) electromagnetic calorimeter (EMC), a superconducting solenoid providing a 1.0 T magnetic field, and a muon counter. The charged particle momentum resolution is $0.5\%$ at a transverse momentum of 1 $\gevc$ and the photon energy resolution in the EMC is $2.5\%$ ($5\%$) in the barrel (endcap) region for 1 $\gev$ photons. A more detailed description of the BESIII detector is given in Ref.~\cite{2009MAblikimDet}.

  High-statistics $\ee$ annihilation Monte Carlo (MC) samples, generated by the {\scshape{geant4}}-based~\cite{2003GeantNIM, 2006GeantIEEE} $\mbox{MC}$ simulation package {\scshape{boost}}~\cite{2006ZYDeng},
  are used to investigate the backgrounds, to optimize the selection criteria, and to determine the detection efficiencies.
  The $\ee$ annihilation is simulated by the $\mbox{MC}$ generator {\scshape{kkmc}}~\cite{2001SJadach}, taking into consideration the spread of the beam energy and the effect of the initial-state radiation (ISR).
  Inclusive $\mbox{MC}$ samples, consisting of $\lambdacp\lambdacm$ events, charmed meson $D_{(s)}^{(*)}$ pair production, ISR returns to lower mass charmonium(-like) $\psi$ states, and continuum QED processes $\ee\to\qqbar$ ($q = u, d, s$), are used to study the backgrounds.
  All known decay modes are generated with $\uchyph=0${\scshape{evtgen}}~\cite{2008RGPing,2001DJLange} with BFs being the values of the Particle Data Group (PDG)~\cite{2016PDG}, and the remaining unknown decay modes are generated by $\mbox{{\scshape{lundcharm}}}$~\cite{2000JCChen}. The signal MC samples of $\ee\to\lambdacp\lambdacm$ are produced with one $\lambdac$ decaying to the final states of interest, $p\eta$ or $p\piz$, and the other $\lambdac$ decaying generically to any of the possible final states.

  Charged tracks, reconstructed from hits in the MDC, are required to have a polar angle $\theta$ satisfying $|\cos\theta|<$ 0.93 and a point of closest approach to the interaction point within $\pm$10 cm along the beam direction ($\vz$) and 1 cm in the plane perpendicular to the beam ($\vr$).
  Information from the TOF is combined with the ionization energy loss ($dE/dx$) from the MDC to calculate particle identification (PID) confidence levels (C.L.) for the $\pi$, $K$, and $p$ hypotheses.
  The mass hypothesis with the highest PID C.L. is assigned to each track.
  A further requirement ${\vr}<$ 0.2 $\rm{cm}$ is imposed on the proton candidates to avoid backgrounds from beam interactions with residual gas inside the beam pipe and materials of beam pipe and MDC inner wall.
  Photon candidates are reconstructed by clustering energy deposits in the EMC crystals.
  Good photon candidates are required to have energies larger than 25 $\mev$ in the barrel region ($|\costht| <$ 0.8) or 50 $\mev$ in the endcap region (0.86 $< |\costht| <$ 0.92).
  To eliminate showers produced by charged particles, showers are required to be separated by more than $20^\circ$ from anti-protons, and by more than $8^\circ$ from other charged particles.
  The EMC time is required to be within (0, 700) ${\rm{ns}}$ of the event start time to suppress electronic noise and showers unrelated to the event~\cite{2015MAblikimLambdac}.
  The EMC shower shape variables are used to distinguish photons from anti-neutrons: the photon candidates are required to have a lateral moment~\cite{lateralmoment} less than 0.4, and $E_{3\times3}/E_{5\times5}$ larger than 0.85, where the $E_{3\times3}$ ($E_{5\times5}$) is the shower energies summed over $3\times3$ ($5\times5$) crystals around the center of the shower.

  In the studies of $\lambdacp\to p\eta$ and $\lambdacp\to p\piz$ decays, the $\eta$ $\righthyphenmin=2$mesons are reconstructed in their two most prominent decay modes, $\eta\to\gamma\gamma$ ($\eta_{\gamma\gamma}$) and $\eta\to\pppm\piz$ ($\eta_{\pppm\piz}$), while the $\piz$ meson is reconstructed in its dominant decay $\righthyphenmin=2$mode $\piz\to\gamma\gamma$.
  Candidate $\eta\to\gamma\gamma$ and $\piz\to\gamma\gamma$ decays are selected using all $\gamma\gamma$ combinations with an invariant mass within 3 times the mass resolution (10 (6) $\mevcc$ for the $\eta$ ($\piz$) signal) of their nominal masses ($M_{\eta}$ or $M_{\piz}$)~\cite{2016PDG}.
  An additional requirement, $|\cos \theta_{\rm{decay}}|<$ 0.9, where $\theta_{\rm{decay}}$ is the polar angle of one $\gamma$ in the helicity frame of the $\gamma\gamma$ system,
  is imposed on the candidate $\eta\to\gamma\gamma$ decay to suppress combinatorial backgrounds.
  To improve the momentum resolution, the $\gamma\gamma$ invariant mass is then constrained to $M_{\eta}$ or $M_{\piz}$ mass, and the resultant momenta are used in the subsequent analysis.
  The candidate $\eta\to\pppm\piz$ are reconstructed using all $\pipi\piz$ combinations with an invariant mass satisfying $|M_{\pppm\piz}-M_{\eta}|<12 \mevcc$.

  The $\lambdacp$ is reconstructed using all combinations of the selected proton and the $\eta(\piz$) candidates. For $\ee$ annihilation at $\sqrt{s}=$ 4.6 $\gev$, there are no additional hadrons produced with the $\lambdacp\lambdacm$ pair due to the limited phase space.
  Thus, two kinematic variables, the beam energy constrained mass $M_{\rm{BC}}\equiv \sqrt{E_{\rm{beam}}^2/c^4-|\overrightarrow{p}_{\lambdacp}|^2/c^2}$ and the energy difference $\Delta E\equiv E_{\lambdacp} - E_{\rm{beam}}$, are used to identify $\lambdacp$ candidates.
  Here, $\overrightarrow{p}_{\lambdacp}$ and $E_{\lambdacp}$ are the reconstructed momentum and energy of the $\lambdacp$ candidate in the $\ee$ c.m.\ system, and $E_{\rm{beam}}$ is the energy of the electron and positron beams.
  For a $\lambdacp$ candidate that is reconstructed correctly, $M_{\rm{BC}}$ and $\Delta E$ are expected to be consistent with the $\lambdacp$ nominal mass and zero, respectively.
  A $\lambdacp$ candidate is accepted if the corresponding $|\Delta E|$ is less than 2.5 times its resolution ($\sigma_{\Delta E}$).
  The decay mode dependent $\Delta E$ requirements, are summarized in Table~\ref{tab:SigBkg0}.
  For a given decay mode, we accept at most one charmed baryon candidate per event, retaining the one with the minimum $|\Delta E|$. If there are candidates from different decay modes, we keep them all. 
  For the decay mode $\lambdacp\to p\eta_{\pppm\piz}$, the peaking background from the CF decay mode $\lambdacp\to\pppm\Sigma^+(\Sigma^{+}\to p\piz)$ is eliminated by requiring the invariant mass of proton and $\piz$ satisfying $|M_{p\piz}-M_{\Sigma^{+}}|> 0.015 \gevcc$. The MC study shows that the residual peaking backgrounds from $\lambdacp\to\pppm\Sigma^+(\Sigma^{+}\to p\piz)$ and from $\lambdacp\to\Lambda\pip\piz (\Lambda\to p\pim)$ and $\lambdacp\to pK_{S}^0\piz (K_{S}^{0}\to\pppm)$, which have exactly the same final states as the signal, are negligible.

\begin{figure}[htbp]
  \centering
  \mbox{
    \begin{overpic}[width=0.38\textwidth, height=0.55\textwidth]{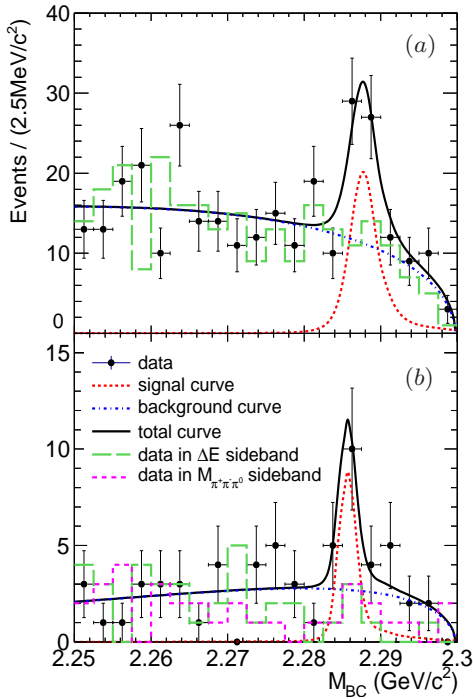}
    \put(58,90){$(a)$}
    \put(58,45){$(b)$}
    \end{overpic}
  }
  \caption{ (color online)
  Simultaneous fit to the $M_{\rm{BC}}$ distributions of $\lambdacp \to p\eta$ reconstructed with the decay modes (a) $\eta\to \gamma\gamma$ and (b) $\eta\to\pppm\piz$.
  The dots with error bars are data, the (black) solid curves are for the best fits, the (blue) dash-dotted curves are for the backgrounds, and the (red) dashed curves are for the signals.
  The (green) long-dashed histograms and (pink) dashed histogram (in (b) only) are the data in the $\Delta E$ and $M_{\pppm\piz}$ sideband region.  }
  \label{fit-eta}
  \end{figure}

\begin{figure}[htbp]
  \centering
  \mbox{
    \begin{overpic}[width=0.38\textwidth, height=0.27\textwidth]{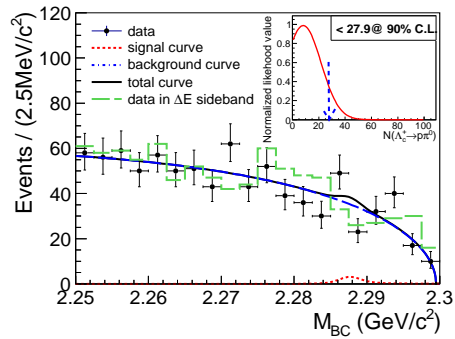}
    \end{overpic}
  }
  \caption{(color online) Fit to the $M_{\rm{BC}}$ distribution for the decay $\lambdacp \to p\piz$. The dots with error bars are data, the (black) solid curve is for the best fit, and the (blue) dashed curve is for the background. The (green) long-dashed histogram is the data in the $\Delta E$ sideband region. The insert shows the normalized likelihood distribution, which includes the systematic uncertainty, as a function of the expected signal yield. The (blue) dashed arrow indicates the upper limit on the signal yield at $90\%$ C.L.}
  \label{fit-pi0}
\end{figure}

  The resultant $M_{\rm{BC}}$ distributions for the decays $\lambdacp\to p\eta$ and $\lambdacp\to p\piz$ are depicted in Fig.~\ref{fit-eta} and Fig.~\ref{fit-pi0}, respectively.
  The $\lambdacp\to p\eta$ signals are seen in both $\eta$ decay modes, but no obvious $\lambdacp\to p\piz$ signal is observed.
  The data in the $\Delta E$ sideband region, defined as 3.5$\sigma_{\Delta E}$ $<|\Delta E|<$ 6$\sigma_{\Delta E}$, are used to study the backgrounds.
  The corresponding $M_{\rm{BC}}$ distributions, illustrated by the long-dashed histograms in Fig.~\ref{fit-eta} and Fig.~\ref{fit-pi0}, show no $\lambdacp$ signal and that the combinatorial backgrounds are well described by the data in the $\Delta E$ sideband region.
  For the decay mode $\lambdacp\to p\eta_{\pppm\piz}$, data in the $\eta$ sideband region ($0.016<|M_{\pppm\piz}-M_{\eta}|<0.032$ $\gevcc$), illustrated by the (pink) dashed histogram in Fig.~\ref{fit-eta}(b), also shows no evidence for peaking background.
  This is further validated by an analysis of the inclusive MC samples, where it is found that the combinatorial backgrounds are dominated by the processes $\epem\to q\bar{q}$.

  To extract the signal yield for the decay $\lambdacp\to p\eta$, we perform unbinned maximum likelihood fits to the $M_{\rm{BC}}$ distributions.
  The signal probability density function (PDF) is constructed by the signal MC simulated shape convoluted with a Gaussian function. Since MC simulation   may be imperfect for modeling of the detector resolution and beam-energy spread of data, the mean and width of Gaussian function are free parameters to account for the potential mass shift and resolution difference between data and MC simulation. The mean ($\mu$) and width ($\sigma$) values of Gaussian function are
  $\mu = (0.74\pm0.56)$ $\mevcc$ and $\sigma = (0.32\pm2.28)$ $\mevcc$
  for $\lambdacp\to p\eta_{\gamma\gamma}$, while $\mu = (-1.22\pm0.80)$
  $\mevcc$ and $\sigma = (0.02\pm1.44)$ $\mevcc$ for $\lambdacp\to
  p\eta_{\pppm\piz}$, respectively.
  The background shape is modeled by an ARGUS function~\cite{1990argus} with the fixed high-end cutoff $E_{\rm{beam}}$.
  The reliability of the ARGUS function is validated with the data in the $\Delta E$ sideband region as well as the inclusive MC samples in the signal region.
 In the decay $\lambdacp\to p\eta_{\pppm\piz}$, the peaking backgrounds from the CF decays have been found to be negligible by MC studies, and are not considered in the fit.
  The fits are performed for the two $\eta$ decay modes separately.
  The corresponding BFs are calculated using
 \begin{eqnarray}
  \BR(\lambdacp\to p\eta) = \frac{N_{\rm{sig}}}{2\cdot N_{\lambdacp\lambdacm}\cdot\eff\cdot \BR_{\rm{inter}}},
  \label{for:branching}
 \end{eqnarray}
  where $N_{\rm{sig}}$ is the signal yield determined from the $M_{\rm{BC}}$ fit, $N_{\lambdacp\lambdacm}=(105.9\pm4.8(\rm{stat.})\pm0.5(\rm{syst.}))\times10^{3}$ is the total number of $\lambdacp\lambdacm$ pairs in the data~\cite{2015MAblikimLambdac}, $\eff$ is the detection efficiency estimated by the MC simulation, and $\BR_{\rm{inter}}$ is the $\eta$ or $\piz$ decay BF taken from the PDG~\cite{2016PDG}. The factor of 2 in the denominator accounts for the charge conjugation of the $\lambdacp$.
  Table~\ref{tab:SigBkg0} summarizes the signal yields, the statistical significances, estimated by the changes in the likelihood values obtained with and without the $\lambdacp$ signal included, the detection efficiencies, and the resulting BFs.
  The two BFs for $\lambdacp\to p\eta$, corresponding to the two $\eta$ decay modes, are consistent within statistical uncertainties.

  \begin{table}[htbp]
  \begin{center}
  \footnotesize
  \caption{Summary of the $\Delta E$ signal regions, the signal $\righthyphenmin=2$yields, the statistical significances, the detection efficiencies, and the BFs (where the first uncertainties are statistical, and the second systematic) for the different $\lambdacp$ decay modes.}
  \begin{tabular}{l c c c c c}
  \hline\hline
                       &  $p\eta_{\gamma\gamma}$   &  $p\eta_{\pppm\piz}$  & $p\piz$            \\ \hline
  $\Delta E$ ($\gev$)  &  $[-0.034, 0.030]$        &  $[-0.027, 0.018]$      &  $[-0.056, 0.029]$ \\
  $N_{\rm{sig}}$       &   $38 \pm 11$             &    $14 \pm 5$           &   $ < 27.9$        \\
  Significance         &    $3.2\sigma$            &    $2.7\sigma$          &   ...         \\
  $\varepsilon(\%)$    &   39.8                    &     20.3                &    49.0            \\
  $\BR(\times 10^{-3})$&   $1.15 \pm 0.33\pm0.10$  &  $1.45 \pm 0.52\pm0.15$ &  $ < 0.27$ \\
  \hline\hline
  \end{tabular}
  \label{tab:SigBkg0}
  \end{center}
  \end{table}

  We also perform a simultaneous fit to the $M_{\rm{BC}}$ distributions for the two $\eta$ decay modes, constrained to the same $\BR(\lambdacp\to p\eta)$ and taking into account the different detection efficiencies and decay BFs of $\eta$. The projections of the fit curves are illustrated in Fig.~\ref{fit-eta}.
  In the fit, the likelihood values of the two individual $\eta$ decay modes are calculated as a function of BF, and are smeared by considering the correlated and uncorrelated systematic uncertainties (discussed in detail below) between the two $\eta$ decay modes according to Refs.~\cite{Convery,Stenson2006}.
 The overall likelihood value in the fit is the product of those for the two $\eta$ decay modes.
  The resultant BF is determined to be $\BR (\lambdacp\to p\eta)=(1.24 \pm 0.28({\rm{stat.}}) \pm 0.10(\rm{syst.})) \times 10^{-3}$ with a statistical significance of $4.2\sigma$, where the significance is estimated by the difference of maximum likelihood values for simultaneous fits with and without signal.

  Since no significant $\lambdacp\to p\piz$ signal is observed, an upper limit on the BF is estimated.
  We fit the $M_{\rm{BC}}$ distribution for the candidate $\lambdacp \to p\piz$ events using similar signal and background shapes to those described previously.
  The result of the best fit is shown in Fig.~\ref{fit-pi0}.
  For the signal PDF, the MC shape is convoluted with a Gaussian function with parameters fixed to those obtained in the fit to $\lambdacp \to p\eta_{\gamma\gamma}$ candidates.
  The PDF for the expected signal yield is taken to be the normalized likelihood $\mathcal{L}$
  obtained by scanning over the signal yield fixed from zero to a large number, and incorporating systematic uncertainties~\cite{Convery,Stenson2006}, as shown in the inset plot of Fig.~\ref{fit-pi0}.
  The upper limit at the 90\% C.L. on the signal yield is $N^{\rm{up}}=$ 27.9 (shown as the arrow in Fig.~\ref{fit-pi0}), corresponding to $\int_0^{N^{\rm{up}}}\rm\mathcal{L}({\it x})d{\it x}$/$\int_0^{\infty}\rm\mathcal{L}({\it x})d\it{x}$ = 0.9.
  The upper limit at the 90\% C.L. on the BF is calculated with Eq.~(\ref{for:branching}) by substituting $\eta$ with $\piz$ and is reported in Table~\ref{tab:SigBkg0}.


  Several sources of systematic uncertainties are considered in the BF measurements. The uncertainties associated with the efficiencies of the tracking and PID for charged tracks are investigated with the samples $\ee\to 2(\pip\pim)$, $\kpkm\pppm$ and $p\bar{p}\pppm$ from data taken at $\sqrt{s}>$ 4.0 $\gev$, and the corresponding (transverse) momentum weighted values are assigned as the uncertainties. The uncertainties due to the $\vr$ requirement and the veto on the CF peaking background in the decay $\lambdacp\to p\eta_{\pppm\piz}$ are investigated by repeating the analysis with alternative requirements ($V_{r}$ $<$ $0.25$ cm and $|M_{p\piz}-M_{\Sigma^{+}}|$ $>$ $0.020$ $\gevcc$). The resultant differences of the BFs are taken as the systematic uncertainties.
  The $\piz$ reconstruction efficiency, including the photon detection efficiency, is studied using a control sample of $D^{0} \to K^{-}\pip\piz$ events from a data sample taken at $\sqrt{s}=$ 3.773 $\gev$. The momentum weighted data-MC differences of the $\piz$ reconstruction efficiencies, which are obtained to be $3.3\%$ and $0.8\%$ for $\lambdacp\to p\eta_{\pppm\piz}$ and $\lambdacp\to p\piz$ decays, are considered as the uncertainties. Similarly, the uncertainty for the $\eta_{\gamma\gamma}$ reconstruction efficiency in the decay $\lambdacp\to p\eta_{\gamma\gamma}$ is determined to be $1.0\%$ by assuming the same momentum-dependent data-MC differences as those for $\piz$ candidates.
  The uncertainties associated with the $\eta$ mass window for $\lambdacp\to p\eta_{\pppm\piz}$, the $\cos\theta_{\rm{decay}}$ requirement for $\lambdacp\to p\eta_{\gamma\gamma}$, the $\Delta E$ requirements, and the photon shower requirements are studied using double-tag $D^{+} \to \pip\eta(\piz)$ events.
  The uncertainties from the $M_{\rm{BC}}$ fit for $\lambdacp\to p\eta$ candidates are studied by alternative fits with different signal shapes, background parameters, and fit ranges, and the resultant changes on the BFs are taken as the uncertainties.
  In the determination of the upper limit on the BF of $\lambdacp\to p\piz$ decay, similar alternative fits are investigated, and the one corresponding to the largest upper limit is selected conservatively.
  The uncertainties in the signal MC model arising from the following sources are considered:
  a) the beam energy spread; b) the input cross section line-shape of $\ee\to\lambdacp\lambdacm$ production for $\uchyph=0$ISR; c) the $\lambdacp$ polar angle distribution in the $\ee$ rest frame; d) the different angular momentum between proton and $\eta(\piz)$ candidates.
  The quadratic sum of the resultant $\righthyphenmin=2$differences in the detection efficiencies is taken as the uncertainty.
  The uncertainties of the MC statistics, the total $\lambdacp\lambdacm$ number quoted from Ref.~\cite{2015MAblikimLambdac} and the decay BFs for the intermediate state decays quoted from the PDG~\cite{2016PDG} are also considered.
  The total systematic uncertainties, quadratic sum of the individual ones, are $8.3\%$, $10.2\%$, and $5.2\%$ for $\lambdacp\to p\eta_{\gamma\gamma}$, $p\eta_{\pppm\piz}$ and $p\piz$, respectively.
  The individual systematic uncertainties are summarized in the Table~\ref{tab:uncertainties}.
  
  \begin{table}[htbp]
  \begin{center}
  \footnotesize
  \caption{Summary of the relative systematic uncertainties in percent for $\Lambda_c^+ \to p\eta_{\gamma\gamma}$, $p\eta_{\pi^{+}\pi^{-}\pi^0}$ and $p\pi^0$. The sources tagged with $'*'$ are 100\% correlated between the two $\eta$ decay modes.}
  \begin{tabular}{l c c c}
      \hline \hline
      Sources         &~~~~$p\eta_{\gamma\gamma}$~~~~& $p\eta_{\pi^+\pi^-\pi^0}$ & $p\pi^0$ \\ \hline
      $^*$Tracking for $p$              & 1.3   & 1.3    & 1.3             \\
      $^*$PID for $p$                   & 0.3   & 0.3    & 0.3             \\
      Tracking for $\pi^{+}\pi^{-}$        & ...    & 2.0    & ...              \\
      PID for $\pi^{+}\pi^{-}$             & ...    & 2.0    & ...              \\
      $^*$$V_{r}$ requirement    & 0.2   & 0.2    & 0.2             \\
      CF peaking background veto      & ...   & 1.3    & ...              \\
      $\eta_{\gamma\gamma}/\pi^0$ reconstruction  &1.0     & 3.3    & 0.8  \\
      $M_{\pi^{+}\pi^{-}\pi^0}$ mass window      & ...    & 1.2    & ...              \\
      $\cos\theta_{\rm{decay}}$ requirement      &1.2   & ...  & ...  \\
      $\Delta E$ requirement          & 0.4   & 1.5    & 0.4             \\
      Shower requirement              & 0.8   & 1.9    & 1.7             \\
      $M_{\rm{BC}}$ fit               & 6.5   & 7.1    & ... \\
      Signal MC model                 & 0.7   & 1.2    & 0.8 \\
       MC statistics                  & 0.1  & 0.1    & 0.1 \\
      $^*$$N_{\Lambda_c^+\bar{\Lambda}_c^-}$    & 4.6  & 4.6    & 4.6 \\
      $\mathcal{B}_{\rm{inter}}$             & 0.5   & 1.2    & negligible \\ \hline
      Total                           & 8.3   & 10.2    & 5.2  \\
      \hline\hline
  \end{tabular}
  \label{tab:uncertainties}
  \end{center}
  \end{table}

  In summary, using 567 $\ipb$ of $\ee$ annihilation data taken at a c.m.\ energy of $\sqrt{s}=$ 4.6 $\gev$ with the BESIII detector,  we find the first evidence for the SCS decay $\lambdacp\to p\eta$ with a statistical significance of $4.2\sigma$ and measure its absolute BF to be $\BR(\lambdacp\to p\eta)=(1.24\pm0.28(\rm{stat.})\pm 0.10(\rm{syst.})) \times 10^{-3}$.
  In a search for the SCS decay $\lambdacp\to p\piz$, no obvious signal is observed and an upper limit at the 90\% C.L. on its BF is determined to be $\BR (\lambdacp \to p\piz) < 2.7 \times 10^{-4}$. The corresponding ratio of BFs between the two decays is also calculated to be $\BR(\lambdacp \to p\piz) /\BR (\lambdacp \to p\eta)< 0.24$, where the common uncertainties are cancelled.
  The measured BFs and their ratio are compared to the theoretical predictions from different models, as shown in Table~\ref{tab:compare}.
  Our measured BF of $\lambdacp\to p\eta$ is consistent, within two standard deviations,  with one of predictions in Ref.~\cite{1997KKSharma}, the one that assumes flavor SU(3) symmetry and negative sign for p-wave amplitude of $\lambdacp\to\Xi^{0}K^+$. It is worth noting that our measurement is significantly higher than other's theoretical predictions. 
  The measured upper limit of $\BR (\lambdacp \to p\piz)$ is compatible with the predicted values of most of theoretical models, but is smaller by a factor of 2 than that in Ref.~\cite{2016lvcaidian}.
  Overall, the obtained relatively large value of $\BR(\lambdacp\to p\eta)$ and the trend toward small value of the ratio $\BR(\lambdacp\to p\piz)$/$\BR(\lambdacp\to p\eta)$ will have a significant impact on theoretical calculation and will be helpful to understand the underlying dynamics of charmed baryon decays and to test SU(3) flavor symmetry. Additional experimental data will improve the sensitivity of the measurements and allow a better discrimination between the different models.

  \begin{table}[htbp]
  \begin{center}
  \begin{threeparttable}[b]
  \footnotesize
  \caption{Comparison of measured BFs (in $10^{-3}$) of $\lambdacp\to p\eta$ and $p\piz$ and their ratio to theoretical predictions.}
  \begin{tabular}{l c c c}
  \hline\hline
    &  ~~$\lambdacp\to p\eta$~~   & ~~$\lambdacp \to p\piz$~~   & ~~$\frac{\BR_{\lambdacp \to p\piz}}{\BR_{\lambdacp \to p\eta}}$~~        \\ \hline
   BESIII                                    &$1.24\pm0.29$      &  $<0.27$    &  $<0.24$  \\
   Sharma \textit{et al}~\cite{1997KKSharma} &$0.2\tnote{a}~(1.7\tnote{b}~)$       &  $0.2$   & $1.0\tnote{a}~(0.1\tnote{b}~)$   \\
   Uppal \textit{et al}~\cite{1994TUppal}        &0.3          & 0.1-0.2    & 0.3-0.7     \\
   S. L. Chen \textit{et al}~\cite{2003SLChen}    &...      & 0.11-0.36\tnote{c}    & ...     \\
   Cai-Dian L$\ddot{u}$ \textit{et al}~\cite{2016lvcaidian}    &...          & 0.45    & ...     \\
  \hline\hline
  \end{tabular}
   \begin{tablenotes}
   \item [a] assume positive sign of p-wave amplitude of $\Lambda_c^+\to\Xi^{0}K^+$
      \item [b] assume negative sign of p-wave amplitude of $\Lambda_c^+\to\Xi^{0}K^+$
   \item [c] calculated relying on different values of parameters b and $\alpha$
   \end{tablenotes}
  \label{tab:compare}
  \end{threeparttable}
  \end{center}
  \end{table}

  The BESIII collaboration thanks the staff of BEPCII, the Institute of High Energy Physics (IHEP) computing center and the supercomputing center of University of Science and Technology of China (USTC) for their strong support.
  P.~L.~Li and H.~P.~Peng are grateful to Prof.\ Hai-Yang
  Cheng for enlightening discussions.
  This work is supported in part by National Key Basic Research Program
  of China under Contract No. 2015CB856700; National Natural Science
  Foundation of China (NSFC) under Contracts Nos. 11125525, 11235011,
  11322544, 11335008, 11425524, 11625523, 11635010, 11375170, 11275189, 11475164, 11475169, 11605196, 11605198; the Chinese Academy of
  Sciences (CAS) Large-Scale Scientific Facility Program; the CAS Center
  for Excellence in Particle Physics (CCEPP); the Collaborative
  Innovation Center for Particles and Interactions (CICPI); Joint
  Large-Scale Scientific Facility Funds of the NSFC and CAS under
  Contracts Nos. U1232201, U1332201, U1532257, U1532258, U1532102; CAS under
  Contracts Nos. KJCX2-YW-N29, KJCX2-YW-N45, QYZDJ-SSW-SLH003; 100 Talents Program of CAS;
  National 1000 Talents Program of China; INPAC and Shanghai Key
  Laboratory for Particle Physics and Cosmology; German Research
  Foundation DFG under Contracts Nos. Collaborative Research Center CRC
  1044, FOR 2359; Istituto Nazionale di Fisica Nucleare, Italy;
  Koninklijke Nederlandse Akademie van Wetenschappen (KNAW) under
  Contract No. 530-4CDP03; Ministry of Development of Turkey under
  Contract No. DPT2006K-120470; National Science and Technology fund;
  The Swedish Resarch Council; U. S. Department of Energy under
  Contracts Nos. DE-FG02-05ER41374, DE-SC-0010504, DE-SC-0010118,
  DE-SC-0012069; U.S. National Science Foundation; University of
  Groningen (RuG) and the Helmholtzzentrum fuer Schwerionenforschung
  GmbH (GSI), Darmstadt; WCU Program of National Research Foundation of
  Korea under Contract No. R32-2008-000-10155-0.

\end{document}